\documentclass[preprint,12pt]{elsarticle}

\usepackage{amsmath, amssymb, amsthm}
\usepackage{graphicx}
\usepackage{booktabs}
\usepackage{siunitx}
\usepackage{hyperref}
\usepackage{enumitem}
\usepackage{float}
\usepackage{setspace}
\usepackage{natbib}
\usepackage{subcaption}

\begin{document}
\begin{frontmatter}
\title{Beta-Dependent Gamma Feedback and Endogenous Volatility Amplification in Option Markets}

\author[sigma]{Haoying Dai\corref{cor1}}
\ead{dhy@terpmail.umd.edu}
\cortext[cor1]{Corresponding author.}

\affiliation[sigma]{
  organization={Eight Sigma Research},
  city={College Park},
  state={MD},
  postcode={20740},
  country={USA}
}
\begin{abstract}
We develop a theoretical framework that aims to link micro-level option hedging and stock-specific factor exposure with macro-level market turbulence and explain endogenous volatility amplification during gamma-squeeze events.
By explicitly modeling market-maker delta-neutral hedging and incorporating beta-dependent volatility normalization, we derive a stability condition that characterizes the onset of a gamma-squeeze event. 
The model captures a nonlinear recursive feedback loop between market-maker hedging and price movements and the resulting self-reinforcing dynamics. 
From a complex-systems perspective, the dynamics represent a bounded nonlinear response in which effective gain depends jointly on beta-normalized shock perception and gamma-scaled sensitivity. 
Our analysis highlights that low-beta stocks exhibit disproportionately strong feedback even for modest absolute price movements. 

\end{abstract}

\begin{keyword}
Gamma squeeze \sep Delta-neutral hedging \sep Endogenous volatility \sep Beta normalization \sep Market microstructure \sep Nonlinear recursive feedback 
\end{keyword}

\end{frontmatter}
\section{Introduction}
Financial markets frequently exhibit nonlinear price dynamics that appear disproportionate to underlying shocks, reflecting the complex interplay between derivative trading, market microstructure, and investor behavior. A salient manifestation of this recursive feedback behavior is the gamma squeeze, in which large option positions compel market makers to adjust their holdings of the underlying asset to maintain delta-neutrality. These hedging flows can self-reinforce, occasionally producing transient deviations far exceed typical volatility. Recent retail-driven episodes, notably in GameStop (GME) and AMC, underscore the practical significance of this nonlinear amplification and its potential implications for market stability\citep{chaumont2021gamestop,lyocsa2022yolo}.

Prior research has established that dealer hedging can generate endogenous price impact. \citep{garleanu2007, garleanu2011} show that option-market liquidity and hedging flows materially affect price formation, while \citet{brunnermeier2009} examine how feedback trading amplifies volatility under varying liquidity conditions. Complementary studies on factor models and systemic risk \citep{adrian2010, engle2002, longin2001} highlight the role of stock-specific sensitivities in propagating volatility, yet most formulations treat feedback as approximately linear and assume constant sensitivity. A broader literature further shows that dynamic hedging, order-flow imbalance, and liquidity fluctuations can produce nonlinear amplification \citep{avellaneda2003, farmer2013, bouchaud2018, cont2014, donier2015}, and that concentrated options positions may trigger abrupt rallies through mechanical gamma rebalancing \citep{ni2008}, consistent with empirical evidence from the 2021 meme-stock events \citep{fisch2022gamestop, huffman2025rise,aggarwal2022meme}. Insights from factor pricing \citep{ang2006cross, campbell1997econometrics, andersen2007, barunik2018} and from nonlinear and critical-transition dynamics \citep{sornette2003, mantegna2000, lux1999, guckenheimer1983, ott2002, kantz2004, farmer2002, lorenz1963, mandelbrot1977, bak1987, bak1988} similarly emphasize how small disturbances can escalate when interacting with constrained liquidity. However, these frameworks still overlook a central asymmetry: feedback intensity depends strongly on how unusual a price change appears relative to a stock’s typical volatility regime.

Empirical observation suggests that identical absolute price changes can elicit markedly different reactions across stocks. For example, a $3\%$ move in a low-beta stock may represent a deviation several times larger than its typical volatility, attracting a large volumes of call options that prompting aggressive hedging adjustments, whereas a $10\%$ move in a high-beta stock may constitute a routine fluctuation. Such nonlinear sensitivity implies that the same exogenous perturbation can trigger vastly different feedback intensities depending on local volatility norms. Hence, it is not appropriate to treat stocks with different $\beta$ levels equivalently when analyzing these events. Capturing this perceived shock intensity relative to stock-specific volatility is essential for understanding differential gamma feedback and the conditions under which a squeeze occurs.

This paper introduces a beta-dependent gamma-feedback framework that integrates delta-gamma hedging mechanics with stock-specific volatility scaling. It considers the relative sensitivity of a specific stock to broader market behavior. The model defines a relative shock measure normalized by $\beta$, enabling a rigorous characterization of conditions under which hedging flows can generate nonlinear amplification. It provides both a mathematically tractable stability condition for gamma squeezes and an intuitive interpretation of why low-beta stocks are disproportionately sensitive to modest shocks.

The main contributions of this work are:
\begin{enumerate}[noitemsep, topsep=0pt]
    \item \textbf{Formal derivation of gamma-squeeze thresholds:} Establishing a condition for endogenous volatility amplification grounded in delta-neutral hedging mechanics.
    \item \textbf{Integration of beta-normalized shock perception:} Providing a framework to study the different behaviors of stocks with varying sensitive to the market turbulence, i.e., different $\beta$s.
    \item \textbf{Simulation and empirical applicability:} The numerical frameworks can be used to synthetic and real market data.
    \item \textbf{A trackable model that can be interpreted to understand market mechanism:} Providing an interpretable model that links option microstructure, market-maker behavior, and factor-driven stock dynamics.
\end{enumerate}

\section{The Model}
\label{sec_2}
\subsection{Formal Gamma-Squeeze Derivation via Delta-Neutral Hedging}
\label{sec_2_1}


The gamma squeeze can be viewed as a recursive feedback system arising from a market maker’s effort to maintain delta-neutrality. When a market maker sells \(N\) call options on a stock with price \(S_t\), they are exposed to changes in delta. Let \(C(S_t,t)\) denote the option price. The delta of a single option is
\begin{equation}
\Delta_c = \frac{\partial C}{\partial S},
\end{equation}
and the total option delta is
\begin{equation}
\Delta_{\mathrm{total}} = N \Delta_c.
\end{equation}
To remain delta-neutral the market maker holds a stock position \(H_t\) such that
\begin{equation}
\Delta_{\mathrm{total}} - H_t = 0 \quad \Rightarrow \quad H_t = N \Delta_c.
\end{equation}

\vspace{0.5ex}
When the underlying price moves, the option's delta changes; market makers must buy or sell the underlying to restore neutrality. This dynamic creates a \emph{feedback loop}: as market makers buy to hedge short-call exposure, buying pressure pushes the price up, which increases option deltas further and forces additional hedging purchases. 

Thus, the sensitivity of delta to changes in the underlying price can be captured: 
\begin{equation}
\Gamma = \frac{\partial \Delta_c}{\partial S} = \frac{\partial^2 C}{\partial S^2}.
\end{equation}
For a small price change \(\Delta S\) the option delta changes approximately as
\begin{equation}
\Delta \Delta_c \approx \Gamma \, \Delta S.
\end{equation}
Consequently, to remain delta-neutral the market maker adjusts their stock holdings by
\begin{equation}
\Delta H_t = N \Gamma \, \Delta S.
\end{equation}

\vspace{0.5ex}

Empirical microstructure literature models the impact of net order flow on prices with a (first-order) linear price–impact coefficient \(\lambda\) \citep{cetin2004liquidity, garleanu2013dynamic}:
\begin{equation}
\Delta S_t = \mu S_t + \lambda \, \Delta H_t,
\end{equation}
where \(\mu S_t\) denotes exogenous drift or shock that ignites a price change and $\mu$ measures the level of surprise and \(\lambda\) summarizes liquidity/price–impact properties of the market. Substituting the hedging adjustment \(\Delta H_t = N \Gamma \, \Delta S_t\) yields
\begin{equation}
\label{eq:Delta_St}
\Delta S_t = \mu S_t + \lambda N \Gamma \, \Delta S_t \quad \Rightarrow \quad \Delta S_t(1 - \lambda G) = \mu S_t,
\end{equation}
where we define total gamma exposure \(G \equiv N \Gamma\).

\subsection{Beta-Normalized Intuition}
\label{sec_2_2}
The derivation in Section~\ref{sec_2_1} captures the mechanical structure of gamma-induced feedback under delta-neutral hedging. However, it implicitly assumes that all price changes of equal absolute magnitude are perceived equally by market participants. In reality, the \textit{sharpness} of a move is not an absolute quantity but a relative one: it depends on how unusual the move appears relative to the stock’s typical behavior and volatility regime.

To formalize this intuition, we introduce a \textit{beta-normalized relative surprise} variable for a specific stock $i$:
\begin{equation}
x_i = \frac{|\Delta S / S|}{|\beta_i| \sigma_m},
\end{equation}
where 
$\sigma_m$ is aggregate market volatility, and $\beta$ measures the stock’s sensitivity to systematic market movements. The $\sigma_m$ captures behavior of the whole market with the implications that during a market-wide high-turbulent event, the stock-specific shocks will be perceived as relatively small and less likely to induce aggressive hedging. Moreover, by normalizing the absolute price change ratio $\Delta S/S$ with respect to $\beta$, the perceived surprise is related to its normal price behavior, justifying the idea the high-volatile stock is less likely to incur gamma squeeze comparing to low-beta stock with the same ratio of price change. 
Here, we focus on the positive beta stocks in a gamma-squeeze event, which leads to $|\beta|= \beta$. This normalization follows the logic of factor-based volatility scaling in asset pricing and systemic risk models \citep{engle2002, longin2001, adrian2019vulnerable}, where $\beta$ effectively acts as a proportionality constant linking market-level shocks to firm-level responses.

\subsection{ Combined Gamma-Beta Model and Proposition}
\label{sec_2_3}
Building on Sections~\ref{sec_2_1} and~\ref{sec_2_2}, we integrate delta-neutral hedging mechanics with beta-normalized shock perception to construct a unified framework for gamma squeezes.

Let $G = N \Gamma$ denote total gamma exposure, 
To remain delta-neutral, the hedging intensity of market makers can be written as:
\begin{equation}
\label{eq:static_hedge}
\Delta H_t = G \, \Delta S_t \, \phi(x), \quad \phi'(x) > 0.
\end{equation}
Here, we introduce $\phi(x)$ as an increasing function that describes how abnormal price movements intensify the scale of hedging adjustments that is directly proportional to the surprise $x_i$ with $\phi(x) = 1 + kx$, with  $k=2$. This form is consistent with both behavioral and empirical findings that order flow becomes more elastic and reactive under unusual price dynamics \citep{bouchaud2018}.

Here, the stability denominator of this stock can be rewritten as follows:
\begin{equation}
\label{eq:Di_def}
D_i = 1 - \lambda \, G \, \phi(x_i).
\end{equation}

From above derivation, it can be seen that gamma squeeze occurs when the stability denominator approaches zero:
\begin{equation}
\label{eq:denominator}
D_i \longrightarrow 0. 
\end{equation}

The condition $D_i \to 0$ formalizes the threshold at which even modest exogenous shocks ($\mu S_t$) are magnified into large price movements.  

Sections~\ref{sec_2_1} to~\ref{sec_2_3} together establish a static theoretical framework for delta-neutral hedging with beta-normalized shock perception, which explains how endogenous volatility amplification can initiate a gamma squeeze event under specific initial conditions. However, a gamma squeeze is inherently a dynamic phenomenon, characterized by a self-reinforcing feedback loop: as market makers hedge by buying the underlying stock, their actions drive the price higher, which in turn attracts additional call-option activity and further intensifies the hedging demand. This dynamic interaction motivates the development of a time-evolving model, which will be discussed in the subsequent section.

\subsection{Dynamic Beta-Normalized Recursive Feedback with Decay and Saturation}
\label{sec_2_4}
In real markets, gamma squeezes evolve through multiple recursive rounds of adjustment, where hedging activity itself becomes the primary driver of subsequent price movements. To capture this compounding mechanism, we formulate a discrete-time recursive model that integrates three interacting components: (i) \textit{position decay}, representing the gradual reduction of dealer exposure as balance-sheet constraints, liquidity costs, or margin limits force contraction of open option positions; (ii) \textit{saturating hedging impact}, which bounds the effective price response through nonlinear liquidity effects; (iii), a decaying exogenous shock term, $\mu_t S_t$, which models the transient initiation of the squeeze and diminishes with the reduction in active hedging exposure.

Let $S_t$ denote the stock price at discrete time $t$, and let market-makers maintain delta-neutrality by recursively adjusting their positions in response to price changes. 
We can replace the static delta neutral hedging Equation~\ref{eq:static_hedge} into a recursive one:
\begin{equation}
\Delta H_t = N_t \Gamma_t \, \Delta S_t^{\text{obs}},
\end{equation}
where $\Delta S_t^{\text{obs}} = S_t - S_{t-1}$ is the observed price change, $N_t$ the active position size, and $\Gamma_t$ the option gamma exposure.

To capture the feedback-sensitive erosion of liquidity and dealer balance-sheet capacity, we model position decay as a function of cumulative realized price movement according to:
\begin{equation}
\label{eq:Nt_cumulative}
N_t = \frac{N_0}{1 + \eta \Big(\sum_{\tau=0}^{t-1} \Big|\frac{\Delta S_\tau^{\mathrm{obs}}}{S_\tau}\Big|\Big)^{\xi}}, 
\qquad \eta, \xi > 0,
\end{equation}
where $\Delta S_\tau^{\mathrm{obs}} = S_\tau - S_{\tau-1}$ and the summation accumulates the absolute relative movements of past price changes, $\eta$ and $\xi$ are hyper-parameters accounting the decay. 
At the onset of a squeeze, cumulative movement is small, implying $N_t \approx N_0$ and strong recursive feedback; 
as large realized moves accumulate, the denominator in \eqref{eq:Nt_cumulative} grows, compressing $N_t$ and reducing hedging pressure. 

Market makers operate under finite inventory and hedging capacity, so their price-impact response cannot grow without bound. Classical inventory research\citep{holt1960planning,arrow1951,arrow1951,sargent1978} shows that limited capacity naturally produces smooth, concave, saturating adjustments rather than unbounded linear responses.
Early models \citep{baumol1952,scarf1960,howard1960} therefore used linear impact capped at a fixed threshold:
\begin{equation}
\label{eq:piecewise_linear}
    I_{\max}(y)=\min\!\bigl(I_{\max},\,\max(-I_{\max},\,y)\bigr).
\end{equation}

However, such piecewise linear function 
introduces nondifferentiabilities and arbitrary turning points. They have been widely criticized for generating model artifacts, numerical instability, and analytical complications in dynamical systems analysis\citep{tobin1969,goodwin1951,fleming2006controlled}.

Thus, we propose the $tanh$ to account for the market maker's limited hedging capacity as a replacement of the piecewise linear function:
\begin{equation}
I(y) = \tanh(cy), \qquad c > 0,
\end{equation}
since it is linear for small order imbalances similar to Equation~\ref{eq:piecewise_linear} but gradually saturates when hedging demand grows and balance-sheet exhausted. It is smooth and avoids an artificial hard cap and abrupt changes that aligns with the idea form Avellaneda--Stoikov framework\citep{avellaneda2008}.

With this, the stock price evolution is modeled as
\begin{equation}\label{eq:recursive_feedback_core}
\Delta S_{t+1} = \mu_t S_t + I\!\big(\lambda_t N_t \Gamma_t \phi(x_t)\big)\,\Delta S_t,
\end{equation}
\begin{equation}
S_{t+1} = S_t + \Delta S_{t+1},
\end{equation}
where $\lambda_t$ is the hedging impact coefficient and $\phi(x_t)$ captures beta-adjusted amplification based on the normalized price surprise,
\begin{equation}
x_t = \frac{|\Delta S_t^{\mathrm{obs}}| / S_t}{\beta \sigma_m}.
\end{equation}

The term $\mu_t S_t$ represents exogenous or semi-endogenous shocks to price (e.g., initial option-driven imbalance). Following the initial disturbance, we model $\mu_t$ as proportional to the remaining active hedging pressure:
\begin{equation}
\mu_t = \mu_0 \frac{N_t}{N_0},
\end{equation}
so that new shocks gradually diminish as market-maker activity decays in proportion to realized cumulative movement.

Equation~\eqref{eq:recursive_feedback_core} describes a coupled recursive feedback loop in which price shocks propagate through market-maker adjustments and cumulative movement–driven decay of option exposure. 

\subsection{Stochastic Extension of Position Dynamics}
\label{sec_2_5}
In previous section, we modeled the dynamic response of market maker's delta-neutral hedging with multiple feedback loops to the original surprise and active option size is gradually decayed by time without accounting the new incoming options. Now, we further extend that model with a stochastic imcoing of new options.

We can assume that after the initial large size option that triggers a gamma-squeeze event, a stochastic incoming of $\nu_t$ with the deterministic decaying $N_t$:
\begin{equation}
\bar{N}_t = N_t + \nu_t,
\label{eq:Nbar}
\end{equation}
Empirical evidence shows that option order flow exhibits persistence due to algorithmic execution, 
inventory smoothing, and clustered retail activity \citep{PanPoteshman2006}.

Thus, we can assume that $\nu_t$ evolves as a state-dependent autoregressive process:
\begin{equation}
\nu_{t+1} = \rho \, \nu_t + \sigma_N \, N_t \, \varepsilon_t,
\qquad \varepsilon_t \sim \mathcal{N}(0,1), \; |\rho| < 1.
\label{eq:stochastic_nu}
\end{equation}
The coefficient $\rho$ controls short-term persistence, while $\sigma_N$ governs the volatility of random deviations relative to the current exposure level $N_t$. 

To prevent unreasonable values of opening option size, 
we impose a one–sided censoring rule and define the effective exposure as
\begin{equation}
\bar{N}_t = \min\!\big\{\,\max\{\,\bar{N}_t,\,0\,\},\,\overline{N}\big\}.
\label{eq:effective_N}
\end{equation}
The lower bound enforces the standard no-short-inventory limit, while $\overline{N}$ acts as a conservative upper reference level derived from the stationary dispersion of the AR(1) deviation process:
\begin{equation}
\overline{N} = N_0 + \kappa\,\frac{\sigma_N N_0}{\sqrt{1-\rho^2}},
\qquad \kappa = 8.
\label{eq:Nbar_cap}
\end{equation}

This upper bound
serves as a conservative numerical safeguard that
prevents the AR(1)–driven deviations from generating unbounded trajectories in
the early stages of the simulation.
Varying $\kappa$ within a wide range produces negligible changes in the simulated paths, confirming that the upper bound plays only a technical role in maintaining well-posed recursion and that the cap is effectively non-binding under realistic conditions.
The stochastic hybrid specification allows for temporary expansion of exposure early in the squeeze, while preserving convergence toward stability in later stages. 
As cumulative movement builds and $N_t$ decays, the effective variance collapses, restoring smooth dynamics. 
The model therefore reproduces both the explosive and the self-damping phases of gamma-induced price dislocation within a single, tractable recursive structure.

\label{sec_2_6}

It is worth noting that the proposed stochastic model can be readily adapted to incorporate empirical market data. For example, the synthetic shocks and arrivals used in the simulations can be replaced by realized option-flow metrics or volatility-surface dynamics extracted from option transaction data. The stochastic component $\nu_t$ may be extracted directly from opening new incoming option size. 
Meanwhile, from an econometric perspective, the structural parameters $\lambda_t$, $\Gamma_t$, $\eta$, and $\xi$,
can be estimated or back-fitted using observed returns and option-flow series. 
In sum, this recursive architecture connects theoretical amplification dynamics with data-driven calibration, offering a tractable empirical approach to study how real-world option positioning propagates into price instability. The framework is therefore ready for direct empirical implementation once market-based estimates of $\lambda_t$, $\Gamma_t$, and $N_t$ are available.

\section{Stability and Bifurcation Structure}
\label{sec:stability_bifurcation}

This section connects the gamma-feedback mechanism to the formal stability
condition associated with the denominator $D_i$ in Equation~\ref{eq:Di_def}.  
The singularity $D_i \to 0$ introduced earlier as the onset of a
gamma-squeeze event is shown to coincide with a stability loss in both the
static price–impact map and the recursive feedback dynamics.  
The resulting threshold defines a codimension-one bifurcation surface in
$(\lambda,G,\beta)$-space.

\subsection{Static stability of the price-impact map}
\label{subsec:static_stability}

Under the linear impact specification of Section~\ref{sec_2_3}, substituting
the delta-hedging relation of Equation~\ref{eq:Delta_St} yields the closed-form
response
\begin{equation}
    \Delta S_t = \frac{\mu S_t}{D_i},
    \label{eq:static_response}
\end{equation}
where $D_i=1-\lambda G\phi(x_i)$.  
When $D_i>0$ the response is locally proportional to the initiating shock,
whereas $D_i\to 0$ produces an unbounded linear amplification.  
The static gamma-squeeze threshold therefore corresponds to
\begin{equation}
    D_i = 0
    \quad\Longleftrightarrow\quad
    \lambda G\phi(x_i)=1,
    \label{eq:static_threshold}
\end{equation}
which defines a codimension-one instability boundary.

\subsection{Local stability of the recursive map}
\label{subsec:dynamic_stability}

To relate the static condition to the dynamic model in
Equation~\ref{eq:recursive_feedback_core}, we freeze the slowly varying
quantities $(S_t,N_t,\Gamma_t,\lambda_t)$ at initial values and obtain the
affine one-dimensional map
\begin{equation}
    \Delta S_{t+1}=a+F\Delta S_t,
    \qquad
    a=\mu_0 S_0,
    \quad
    F=I\!\big(\lambda N_0\Gamma_0\phi(x_0)\big),
    \label{eq:affine_map}
\end{equation}
from which the fixed point 
\[
\Delta S^*=\frac{a}{1-F}
\]
is stable when
\begin{equation}
    |F|<1.
    \label{eq:linear_stability_condition}
\end{equation}

In the linear-impact case $I(y)=y$, the feedback factor becomes
$F=\lambda G_0\phi(x_0)$, and~\eqref{eq:linear_stability_condition} reduces to
\begin{equation}
    \lambda G_0\phi(x_0)<1,
    \label{eq:dynamic_condition}
\end{equation}
which is identical to the static threshold~\eqref{eq:static_threshold}.  
Thus the classical gamma-squeeze denominator $D_i$ also governs the
eigenvalue of the price-impact recursion.

\subsection{Fast-subsystem stability and the role of decay}
According Equation~\ref{eq:recursive_feedback_core}, the map $(S_t,\Delta S_t,N_t)\mapsto (S_{t+1},\Delta S_{t+1},N_{t+1})$ is generally non-autonomous, but over short horizons
the slow variables $(S_t,N_t,\Gamma_t)$ evolve much more gradually than $\Delta S_t$.
We therefore analyze the {\it fast subsystem} obtained by treating $(S_t,N_t,\Gamma_t,\lambda_t)$
as quasi-static parameters. 
and macro-finance.
Setting $\Delta S_{t+1}=\Delta S_t=\Delta S^* =0$ we get,

\begin{equation}
\Delta S^*\big(1 - F(\Delta S^*)\big)
=
\mu_t S_t,
\label{eq:fixed_point_simplified}
\end{equation}
where
\[
F(\Delta S)
=
I\!\big(\lambda_t N_t\Gamma_t\,\phi(x)\big).
\]
 
The economically natural equilibrium arises when the initiating shock vanishes,
\[
\mu_t S_t = 0 
\quad\Rightarrow\quad
\Delta S^* = 0.
\]
We now assess stability of the fixed point $\Delta S^*=0$.

Linearizing Equation~\ref{eq:recursive_feedback_core} at $\Delta S_t=0$
gives a local feedback coefficient
\begin{equation}
    F_0 = I\!\big(\lambda_t N_t\Gamma_t\big),
    \label{eq:F0_def}
\end{equation}
The fixed point $\Delta S^*=0$ is locally asymptotically stable if and only if
\[
|F_0| < 1,
\]

since the linearized map is $\Delta S_{t+1}=F_0\Delta S_t$, a one-dimensional linear system with its stability requires $|F_0|<1$.

\subsection{Bifurcation structure}
\label{subsec:bifurcation_structure}

In the linear-impact benchmark $I(y)=cy$ and with fixed exposure $G_0=N_0\Gamma_0$,
the fast subsystem reduces to the affine one-dimensional map.

The equilibrium $\Delta S^* = 0$ (for $\mu_0 S_0=0$) is linearly stable if 
$|F_0|<1$ and loses stability as $F_0$ crosses the unit circle.  
In particular, the fixed point undergoes a classical flip (period-doubling)
bifurcation at $F_0=-1$, while loss of stability relevant for gamma squeezes
occurs as $F_0 \nearrow 1$.

Expressed in terms of the original parameters, the associated threshold is
\[
\lambda G_0\,\phi(x_0,\beta) = 1,
\]
which coincides with the static singularity of the gamma-feedback denominator
$D_i = 1 - \lambda G \phi(x_i)$.  
The corresponding bifurcation surface in parameter space can be written as
\begin{equation}
\mathcal{B}
=
\Big\{(\lambda,G,\beta)\;:\; 1-\lambda G\,\phi(x)=0\Big\},
\end{equation}
separating a weak-feedback region ($D_i>0$, small amplification) from the
strong-feedback regime ($D_i\approx 0$) associated with extreme sensitivity
to shocks.  In this linear benchmark, crossing $\mathcal{B}$ would lead to
unbounded divergence of $\Delta S_t$.

\subsection{Impact of saturation and self-limitation}
\label{subsec:saturation_selflimiting}

The full model replaces linear impact with a saturating response
$I(y)=\tanh(cy)$.  In this case the effective feedback coefficient
\[
F_t = I\!\big(\lambda G_t\,\phi(x_t)\big)
\]
is bounded in magnitude, with
\[
\tanh(cy)\to 1\quad\text{as}\; y\to\infty.
\]
Thus, the strict divergence implied by the linear benchmark for $F_0>1$
is replaced by large but finite excursions: as the system is pushed toward
the instability surface $\mathcal{B}$, the impact function saturates and
$F_t$ remains confined to $(-1,1)$.

At the same time, the effective exposure $G_t = N_t\Gamma_t$ decays
endogenously as cumulative price movement increases.  Early in a squeeze
we have $G_t\approx G_0$ and the dynamics are well approximated by the
linear threshold $D_i=0$.  As the episode unfolds and realized moves
accumulate, $N_t$ shrinks, $G_t$ falls, and the product 
$\lambda G_t\,\phi(x_t)$ is driven back below the instability
boundary.  The combined effect of saturation and exposure decay therefore
produces a self-limiting mechanism: shocks can trigger large transient
amplification when parameters lie near $\mathcal{B}$, but the dynamics
remain bounded and eventually return to the stable regime.

\section{Numerical Simulation and Illustrative Results}
\label{sec_3}

\subsection{Stability Denominator $D_{i}$ and Extreme-Event Visualization}
\label{sec_3_1}
\begin{figure*}[t]
\centering
\begin{tabular}{c c}

\includegraphics[width=6.5cm]{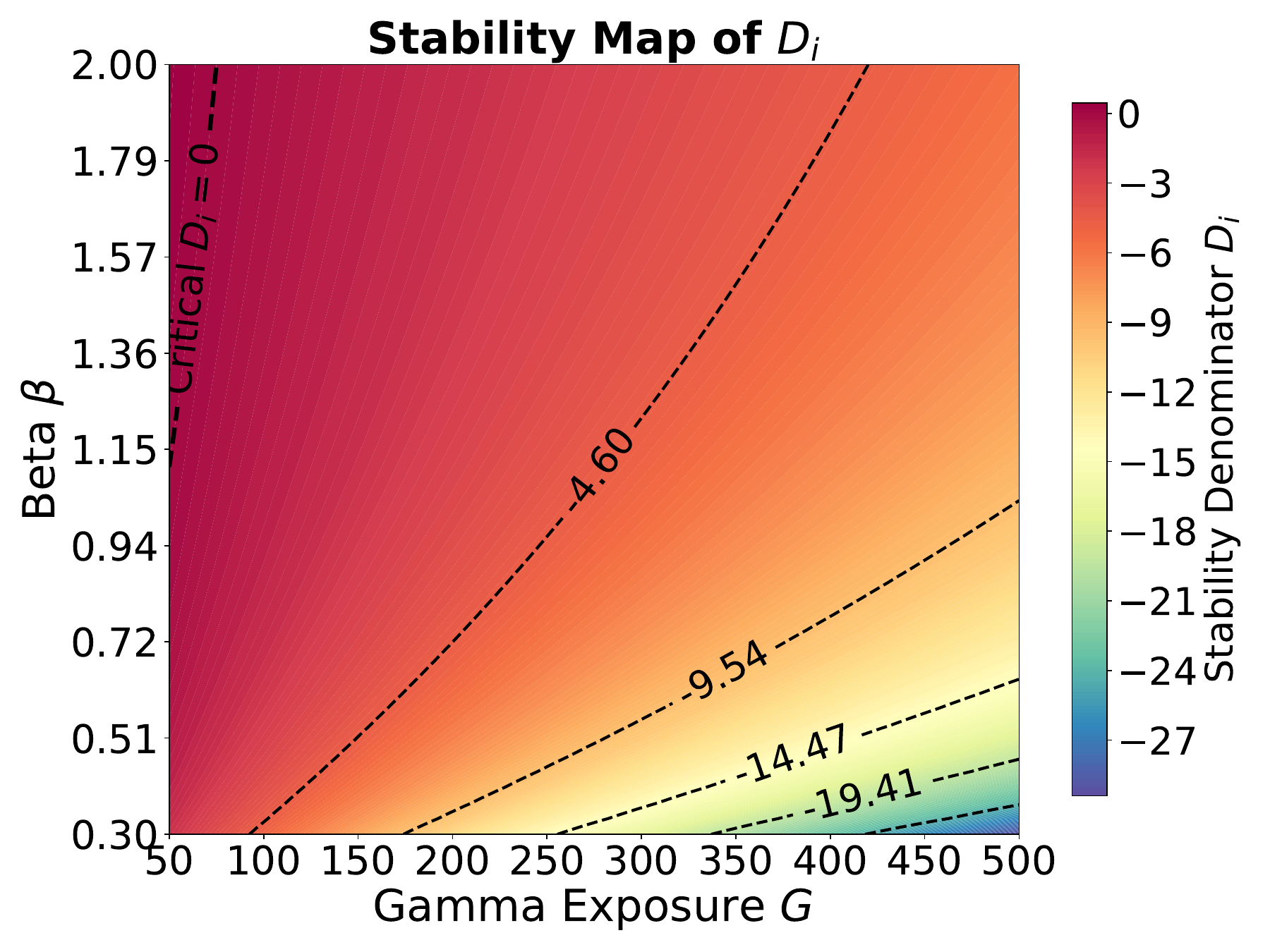} &
\includegraphics[width=6.5cm]{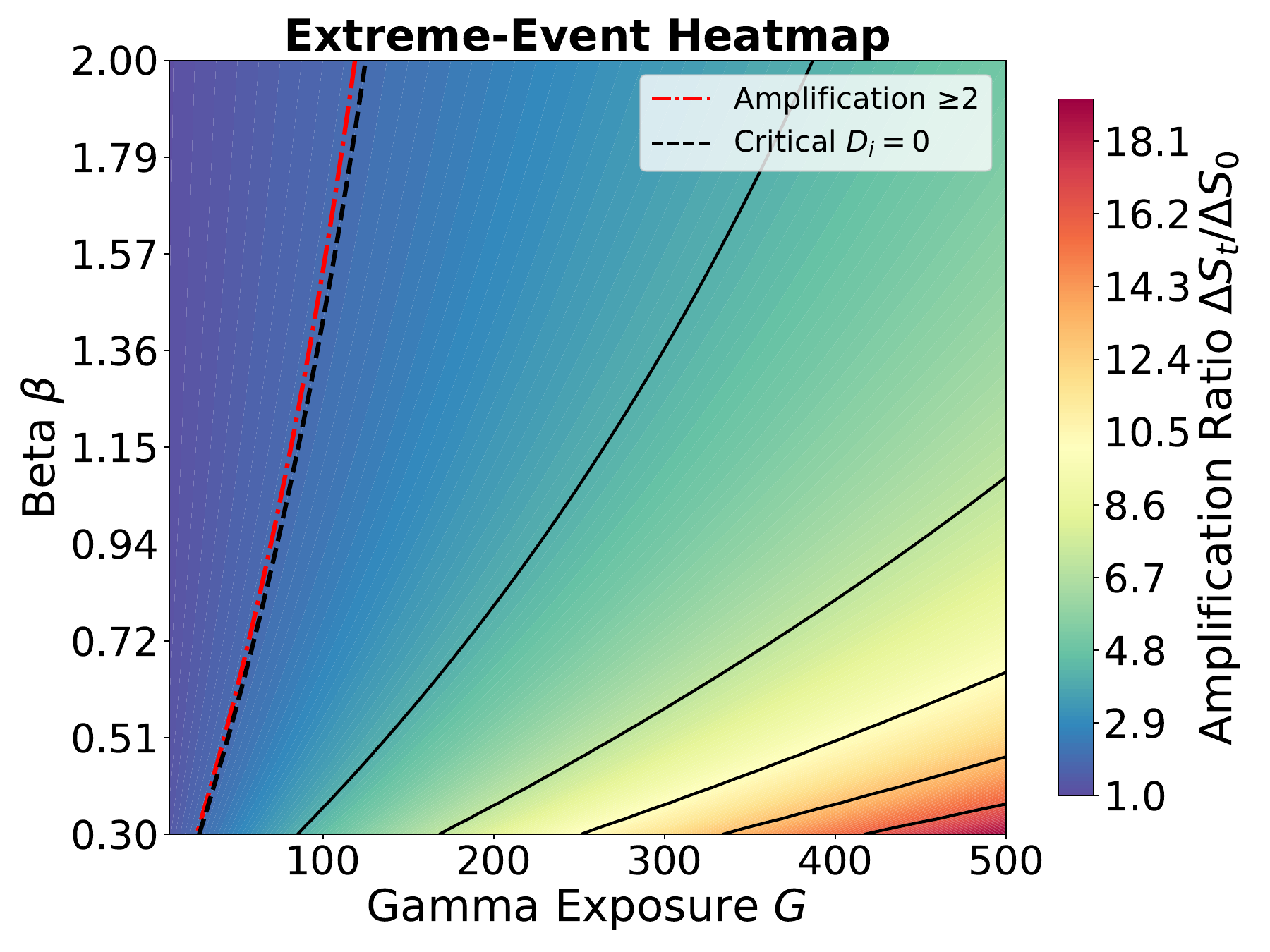} \\
(a) & (b) \\

\end{tabular}

\caption{
\label{fig:stability_heatmap_combined}
(a) Stability map of market-maker feedback dynamics. Color intensity represents the stability denominator $D_i$, with warmer colors corresponding to higher stability. The black dashed line indicates the critical threshold $D_i = 0$, where even small exogenous shocks can produce disproportionately large price movements.
(b) Extreme-event amplification and stability contours. The red dash-dotted line indicates amplification $\ge 2$, while the black dashed line denotes the critical stability boundary $D_i = 0$. These contours reveal regions of pronounced gamma feedback and potential instability.
}

\end{figure*}

To provide a concrete illustration of the beta-dependent gamma feedback model introduced in Section~\ref{sec_2}, we simulate the stability denominator $D_i$ over a finely discretized grid of stock betas ($\beta$) and total gamma exposures ($G$), holding the hedging sensitivity $\lambda$ and the exogenous price shock $\Delta S/S$ fixed. The resulting heatmap, presented in Figure~\ref{fig:stability_heatmap_combined}(a), shows a continuous mapping of $D_i$ across the parameter space, with overlaid contour lines highlighting critical thresholds.
The black dashed line at $D_i=0$ formally delineates the onset of instability, where the market-maker’s delta-hedging can amplify even modest shocks into rapid price escalation. This figure directly illustrates the theoretical stability condition derived in Section~\ref{sec_2_3}, which affirms findings link to the previous model to observable parameter combinations \citep{garleanu2007, brunnermeier2009}. 

It confirms that Stocks with low systematic risk (\(\beta\)) exhibit higher relative surprise for a given absolute shock as the same price move represents a larger deviation relative to typical fluctuations.  Meanwhile, it also reveals total gamma exposure $G$ modulates the magnitude of feedback: larger gamma positions correspond to smaller $D_i$, consistent with stronger delta-hedging-induced amplification \citep{garleanu2011}.

Next, we investigate extreme-event dynamics in the market-maker feedback model, focusing on regions where amplification is particularly pronounced and where stability may be compromised. Specifically, we examine regions in the parameter space where the amplification ratio $\Delta S_t / \Delta S_0$ exceeds 2, highlighting scenarios in which small initial price shocks are significantly magnified due to feedback from gamma hedging. Concurrently, we consider the contour where the stability denominator $D_i$ approaches zero, identifying combinations of parameters at the boundary of instability beyond which the model predicts potential divergence in hedging feedback.

Similarly, we generate a heatmap of amplification across $\beta$ and $\Gamma$ for a representative hedging sensitivity ($\lambda = 0.003$) and shock magnitude ($\Delta S_0 = 0.05$). The heatmap is overlaid with two key contours: a red dash-dotted line marking amplification greater than or equal to 2, and a black dashed line representing $D_i = 0$.

Figure~\ref{fig:stability_heatmap_combined}(b) shows that extreme amplification occurs primarily at low values of $\beta$ combined with moderate-to-high gamma exposure $\Gamma$. This observation aligns with the intuition that low market impact sensitivity and high gamma exposure amplify feedback effects. The $D_i = 0$ contour runs close to the amplification threshold, indicating that scenarios with extreme price amplification are often near the boundary of instability. 

It also reveals that regions of high amplification (e.g., $\Delta S_t / \Delta S_0 \geq 2$) often coincide with or lie near the $D_i = 0$ contour. This alignment underscores the predictive value of the stability denominator in identifying zones of potential market instability. Lastly, the sensitivity of amplification to small variations in $\beta$ or $G$ suggests that even minor changes in market conditions could trigger significant price movements, consistent with observed phenomena in gamma squeezes and other short-term volatility spikes.

\subsection{Time-Series Amplification Dynamics Due to Initial Hedging}
\label{sec_3_4}

In this section, we examine the initial delta-neutral hedging of market-maker to the shock and how that impact the stock price without recursive feedback.
Here, we set the initial price as $S_0 = 100$ such that the stock price movement can be checked later without loss of generality. We set $\lambda = 0.05$, 
$G = 200$ that resembles a common stock in market. 
We compute $\Delta S_t = \mu S_t + \lambda G \Delta S_t \, \phi(x_i)$ iteratively over time as a one-time only hedging reaction. 

\begin{figure}[H]
    \centering
    \includegraphics[width=0.75\textwidth]{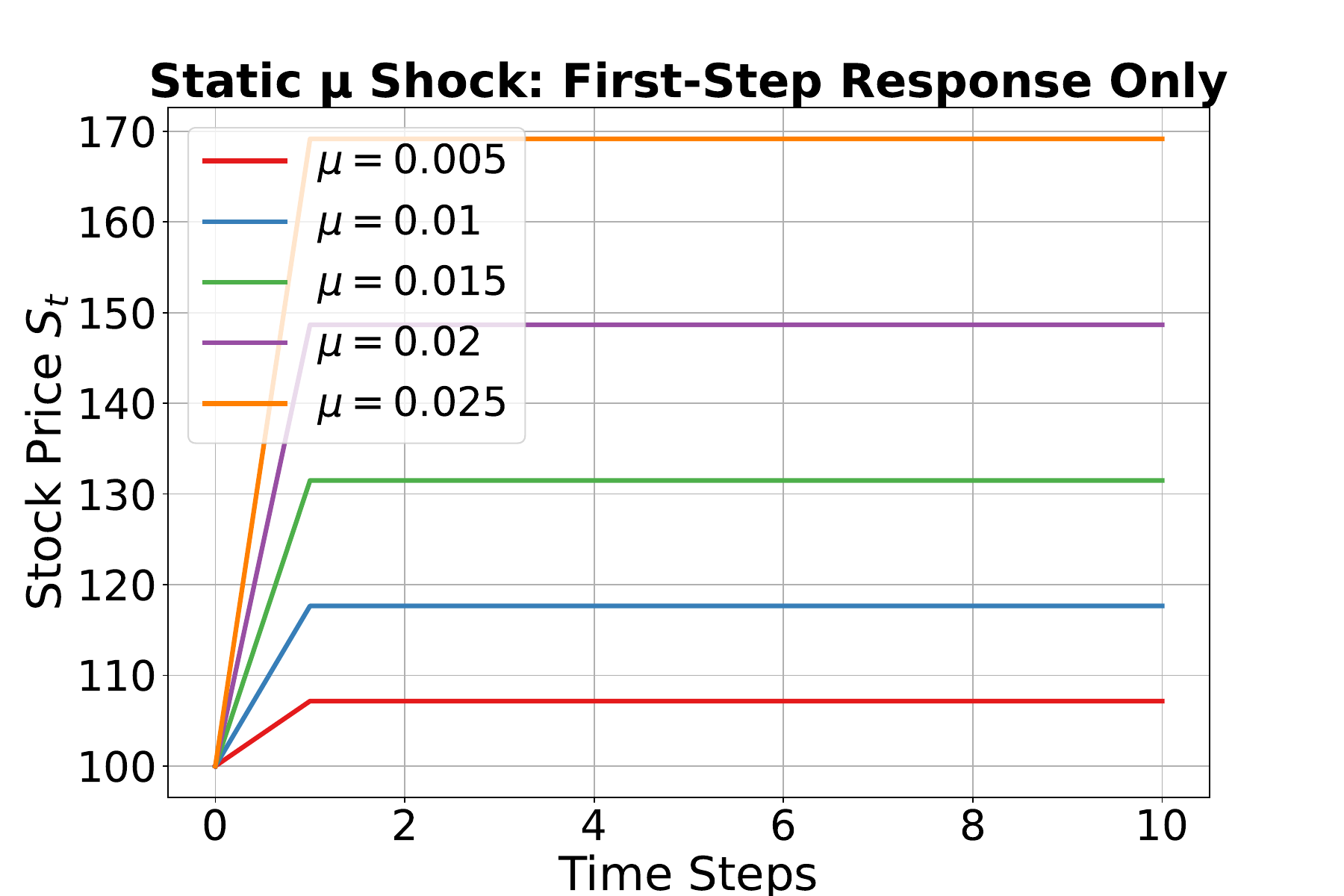}
    \caption{Time-series response to initial $\mu_t$ shocks, using the first-step market impact. 
    Stock price changes are computed according to 
    $\Delta S_t = \mu S_t + \lambda G \Delta S_t \, \phi(x_i)$, applied iteratively to visualize the trajectory. 
    Different curves correspond to different shock values $\mu_t$. 
    }
    \label{fig:amplification_timeseries_mu}
\end{figure}

\begin{figure}[h!]
    \centering
    \includegraphics[width=0.75\textwidth]{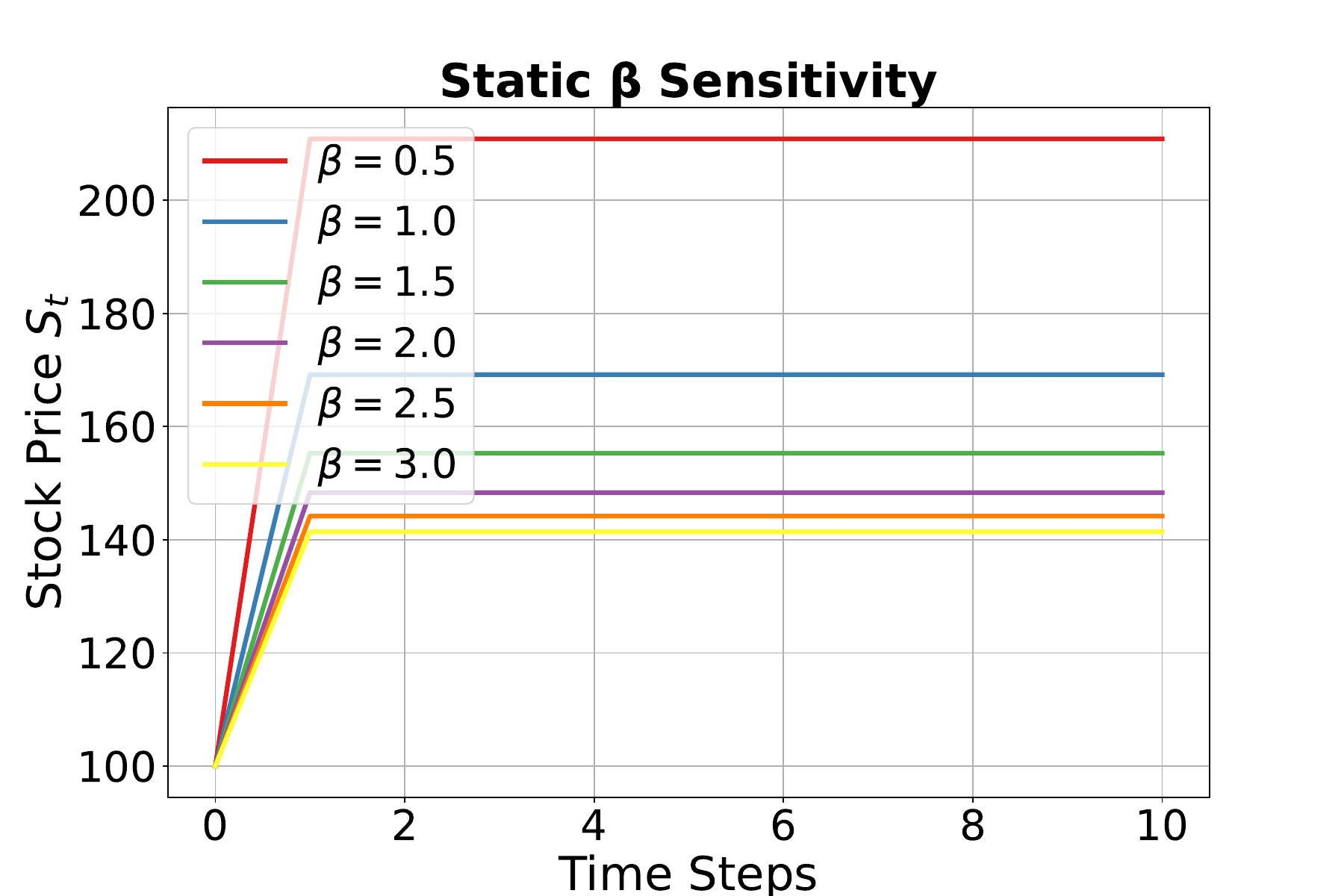}
    \caption{Time-series response to a fixed initial shock $\mu = 0.025$ under
    varying levels of $\beta$.  
    Lower-$\beta$ stocks perceive the same absolute price change as a larger 
    normalized surprise, yielding stronger hedging demand and a larger initial
    price displacement.  The trajectories reflect only the one-time 
    delta-neutral hedge and therefore plateau after the initial jump.}
    \label{fig:amplification_timeseries_beta}
\end{figure}

Figure~\ref{fig:amplification_timeseries_mu} shows the stock price response to 
different levels of initial $\mu_t$ shocks values $\mu_t = 0.005, 0.01, 0.015, 0.02, 0.025$.
It can be seen that a one-time delta-neutral hedging causes a sharp price rise in the underlying stock, followed by a plateau. 
A larger shock leads to a more aggressive hedging, which in turn, increases the price more significantly.
Such behavior is consistent with the notion of a static initial trigger, where the price reacts to the immediate hedging demand but does not yet incorporate the recursive effects that can further amplify the stock movement.
Then, we keep the shock $\mu = 0.025$ and varies $\beta$ from $0.5 - 3$ to check the price behavior of different stocks under the same level of ``surprise''. Figure~\ref{fig:amplification_timeseries_beta} displays the resulting price paths.  
Although each curve corresponds to the same underlying shock, the magnitude of the 
initial price jump varies systematically with $\beta$. It can be seen that the same hedging leads to a shaper price change with respect to stock with lower $\beta$, which is consistent with analysis. 

\subsection{Simulation of Deterministic Recursive Feedback Dynamics} 
\label{sec_3_6}

Building on the dynamic framework introduced in Section~\ref{sec_2_4}, we now conduct numerical simulations to analyze the nonlinear price dynamics arising from recursive hedging feedback.
The market volatility parameter is set to $\sigma_m = 0.03$ which corresponds to the averaged market volatility level. 
\begin{figure}[!ht]
\centering
\includegraphics[width=0.69\textwidth]{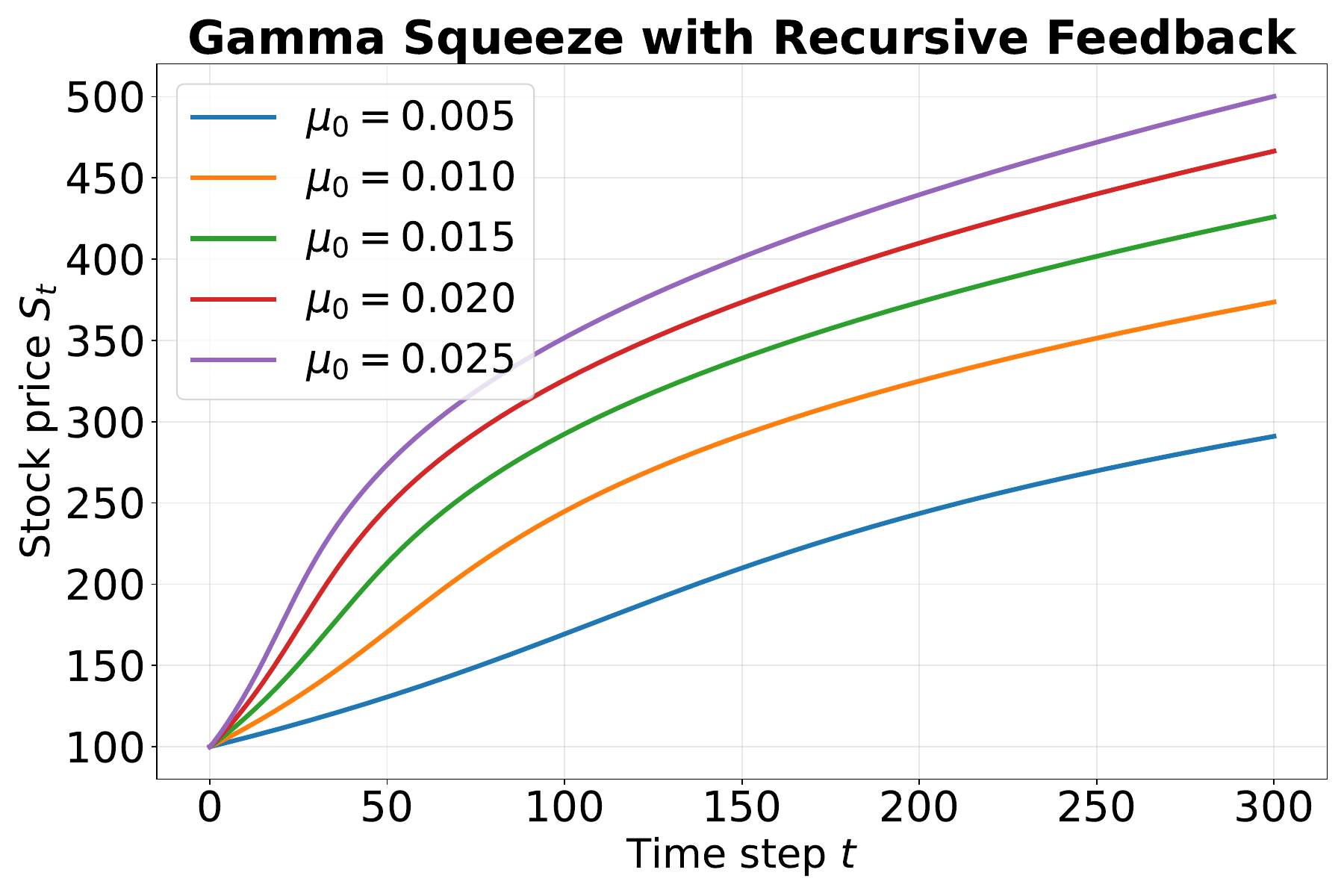}
\caption{Simulated stock price trajectories under recursive feedback with cumulative position decay and saturating impact. 
Larger initial shocks produce steeper early-stage amplification, while cumulative decay of $N_t$ and saturation of $I(\cdot)$ gradually dampen recursive growth, leading to a stable plateau.}
\label{fig:gamma_squeeze_sim}
\end{figure}
First, we check the price response to multiple initial shock magnitudes $\mu_0 \in \{0.005, 0.01, 0.015, 0.02, 0.025\}$ with $\beta =1$, $\eta = 2$, and $\xi=5$, respectively. 
Figure~\ref{fig:gamma_squeeze_sim} illustrates the stock price evolution under multiple exogenous shocks $\mu_t$, demonstrating how recursive hedging amplifies price movements through nonlinear feedback and saturation. 
Early in the process, recursive hedging induces rapid, nonlinear amplification of prices-consistent with the explosive onset of a squeeze event. 
As both $N_t$ and $\mu_t$ decay via cumulative exposure effects, the effective feedback term weakens, and the dynamics transition toward a damped, quasi-stable regime.
The figure also highlights the role of nonlinear saturation and position limits, which prevent unbounded growth even under strong hedging pressure. 

\begin{figure}[!ht]
    \centering
    \includegraphics[width=0.75\textwidth]{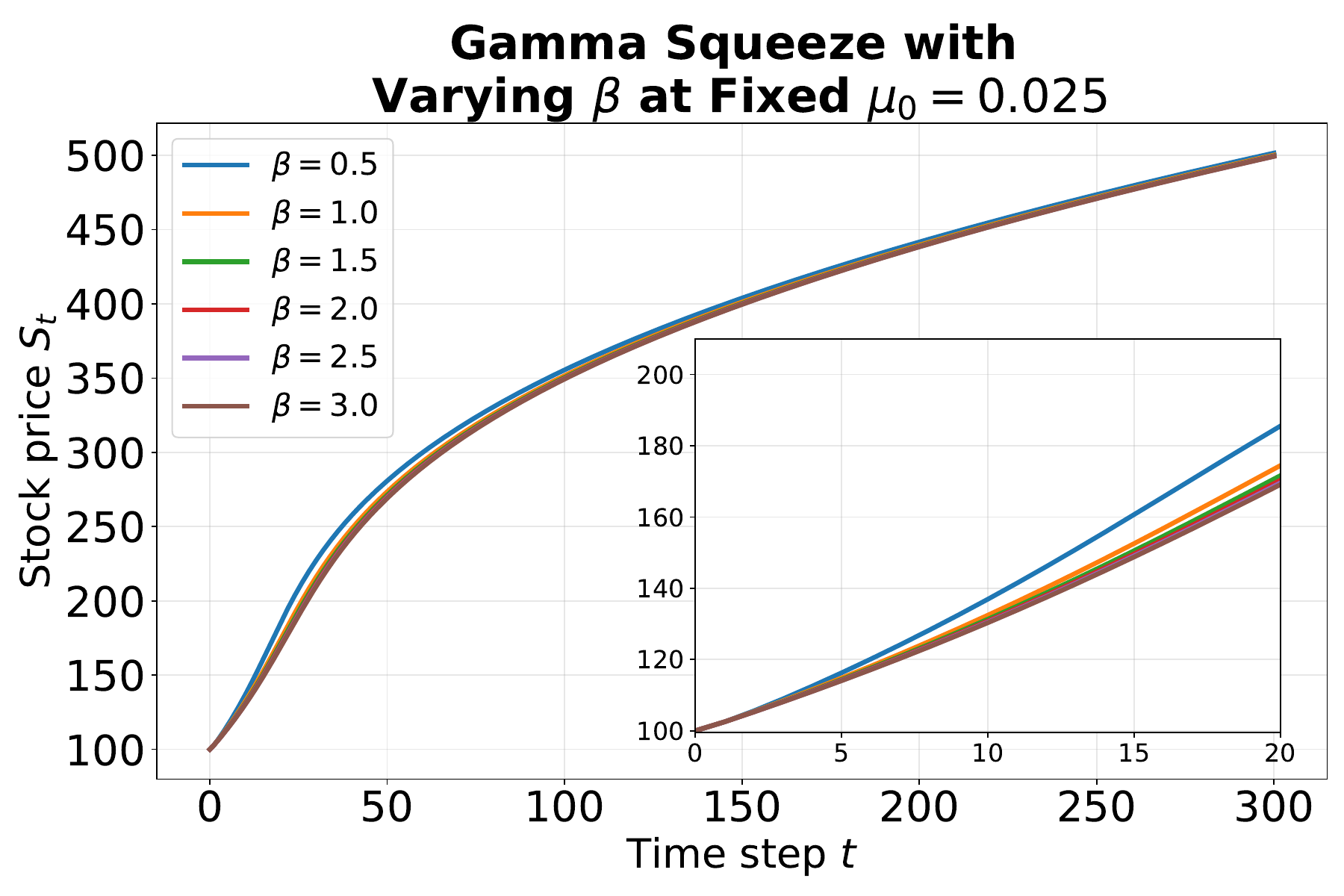}
    \caption{
    Time-series trajectories of stock prices under recursive hedging feedback for varying 
    $\beta$ values, given a fixed initial shock $\mu_0 = 0.025$.  
    Lower-$\beta$ stocks exhibit a stronger initial displacement, as the same absolute shock 
    corresponds to a larger normalized surprise $x_i$, triggering more aggressive hedging.  
    The inset highlights the early-stage divergence, while the long-run trajectories converge 
    as saturation and inventory decay dominate the dynamics.
    }
    \label{fig:gamma_squeeze_multi_beta_with_zoom}
\end{figure}

In contrast to Figure~\ref{fig:amplification_timeseries_mu}, where only single-step hedging was considered, the recursive inclusion of beta-adjusted feedback in this simulation produces sustained, exponential-like growth before eventual stabilization. Figure~\ref{fig:gamma_squeeze_multi_beta_with_zoom} shows that cross-sectional differences in $\beta$ 
produce visible divergence at the onset of the shock: lower-$\beta$ stocks display sharper upward 
movement because an identical surprise constitutes a larger normalized surprise $x_i$, 
leading to stronger initial delta-neutral hedging pressure.  
As the system evolves, however, the recursive feedback loop combined with inventory decay and the 
bounded impact function $I(y)$, progressively dominates the dynamics.  
This behavior provides a structural explanation for why many existing models treat stocks with 
different $\beta$ values as exhibiting similar responses under option-driven feedback: the endogenous market mechanism reacts quickly to overwhelms the early 
cross-sectional divergence. 
However, 
in early stage of gamma-squeeze event, low-beta stocks still experience disproportionately strong amplification before the market mechanism reasserts dominance.

\subsection{Simulation of Recursive Feedback under Stochastic Option Exposure}
\label{sec_3_7}

In this section, we extend our simulation 
for a stochastic model.
Changing the $N_t$ by $\bar{N}_t$ in Equation~\eqref{eq:recursive_feedback_core}, we get:
\begin{equation}
\Delta S_{t+1} = \mu_t S_t + I\!\big(\lambda_t \bar{N}_t \Gamma_t \,\phi(x_t)\big)\,\Delta S_t,
\end{equation}

Firstly, we will check the the a state-dependent autoregressive process of stochastic incoming options in Equation~\ref{eq:stochastic_nu}. Meanwhile, as noted in Section~\ref{sec_2_6}, the recursive framework is designed to accept
empirical option-flow series or volatility-surface data like $\lambda_t$ and $\Gamma_t$ as exogenous drivers.
In the absence of such data, we illustrate this ``plug-in'' capability using a
synthetic example with event-driven option arrivals with random spikes $\nu_t$ 
short-lived amplifications in the price trajectories. Specifically, at randomly selected discrete times $t \in \mathcal{T}_{\text{spikes}}$, 
the exposure receives an additive increment $\nu_t \sim U(0,\,0.3N_0)$ (with $U$ being a uniform distribution), while $\nu_t = 0$ otherwise. All the simulation parameters remain the same as previous section with the stochastic component $\rho=0.9,\ \sigma_N=0.2.$

\begin{figure}[H]
\centering
\includegraphics[width=0.75\textwidth]{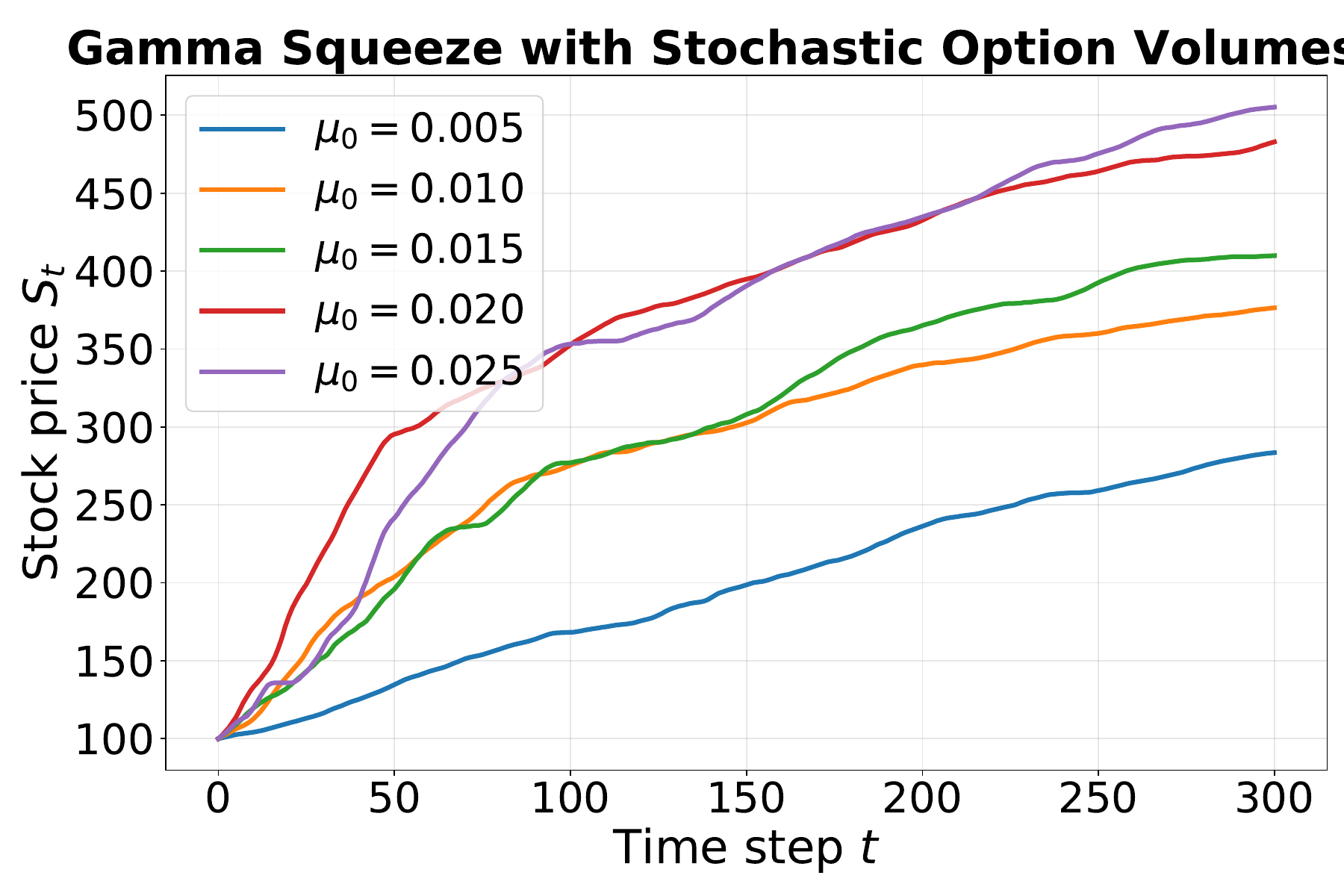}
\caption{Stochastic recursive feedback with cumulative decay. Random deviations in $\bar{N}_t$ induce short-lived reactivations early in the squeeze, while cumulative decay and saturation restore stability.} 
\label{fig:gamma_squeeze_stochastic}
\end{figure}

Figure~\ref{fig:gamma_squeeze_stochastic} shows simulated price trajectories under the stochastic hybrid model. 
Relative to the deterministic case (Figure~\ref{fig:gamma_squeeze_sim}), stochastic deviations in $\bar{N}_t$ produce brief secondary accelerations (“ripples”) early in the squeeze when exposure is still large. 

The stochastic trajectories retain the characteristic two-phase pattern: an initial explosive amplification followed by gradual stabilization. 
However, small fluctuations around the baseline path highlight how the incoming options temporarily impact the underlying stock prices.

\begin{figure}[!ht]
\centering
\includegraphics[width=0.85\textwidth]{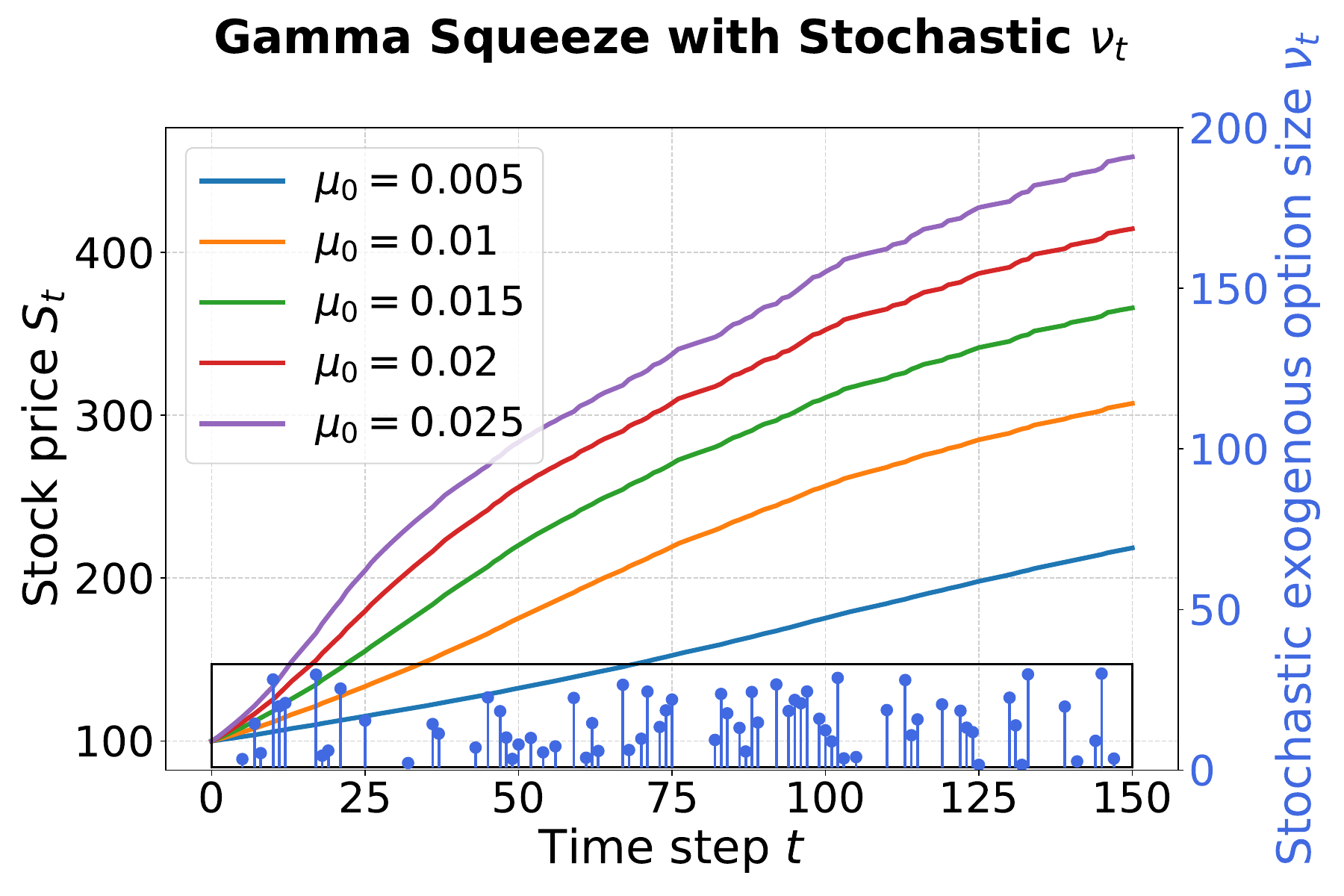}
\caption{Event-driven stochastic extension with synthetic option arrivals $\nu_t$. 
Seventy random spikes $\nu_t \in [0,0.3N_0]$ (blue stems, right axis) generate short-lived amplifications in the price trajectories $S_t$ (left axis). 
All other parameters follow Section~\ref{sec_3_7}.}
\label{fig:gamma_squeeze_discrete}
\end{figure}

Figure~\ref{fig:gamma_squeeze_discrete} shows the simulated price trajectories under synthetic random arrivals $\nu_t$. 
Although each inflow represents at most $30\%$ of $N_0$, these discrete perturbations induce short-lived secondary amplifications superimposed on the decaying price path. 
The nonlinear feedback mechanism magnifies even modest synthetic arrivals, generating visible deviations in $S_t$ analogous to transient ``gamma ignition'' episodes observed in real markets. 

The exercise confirms that the recursive structure can accommodate both continuous and discrete stochastic exposures within a unified analytical framework.
Meanwhile, this controlled extension also confirms that the recursive feedback framework is structurally flexible and empirically adaptable.

\section{Discussion}
\label{sec_4}

A caveat in interpreting the static time-series experiment is that it assumes an instant, unsaturated delta-neutral hedge. This yields an upper-bound ``free response,'' in which the market maker lifts the price without inventory or risk limits. When contrasted with the recursive model, where exposure decays and saturation limits bind, a sharp discrepancy emerges. The unconstrained model produces strong $\beta$-dependent divergence, whereas the recursive, capacity-limited system suppresses much of this cross-sectional spread. Once saturation dominates, feedback no longer scales linearly with $\beta$, and the trajectories for low-$\beta$ and high-$\beta$ names converge.

This contrast highlights an important mechanism. Stronger hedging limits (larger $\eta$, $\xi$, or saturation $c$) reduce the sensitivity of a gamma squeeze to underlying $\beta$ differences. In practical terms, well-capitalized or tightly risk-managed market makers may inadvertently reduce the visible influence of $\beta$ during turbulent periods. From a risk-management perspective, this suggests that robust inventory control can buffer $\beta$-amplified instability and restrict the conditions under which a pronounced gamma squeeze can propagate through the cross-section.

More broadly, this attenuation of cross-sectional dispersion provides a natural explanation for why most existing studies largely overlook $\beta$ effects: in real markets, hedging capacity limits and saturation behavior tend to mask the very $\beta$-dependent spread that the unsaturated model reveals. Consequently, the empirical prominence of $\beta$ is diminished unless hedging frictions are minimal or market stress forces the system into its unlimited regime.

While the present simulations rely on synthetic $\nu_t$ arrivals for clarity, the same recursive framework is immediately adaptable to empirical data. 
As discussed in Section~\ref{sec_2_6}, the stochastic component $\nu_t$ can be replaced with observed option-flow metrics, while $\lambda_t$ and $\Gamma_t$ can be estimated dynamically from market microstructure data. 
Thus, the model is structurally ready for calibration to actual trading environments, bridging theoretical simulation with empirical implementation.

As for limitations of current model, the framework abstracts several real-market complexities. 
First, the price-impact coefficient $\lambda_t$ is modeled in reduced form, whereas empirical impact is known to be state-dependent and asymmetric. 
Moreover, the decay rule for option exposure $N_t$ proxies for balance-sheet contraction but does not explicitly represent volatility targeting, or dealer optimization. 

Finally, $\Gamma_t$ and $\lambda_t$ are treated as exogenous or slow-moving despite evolving endogenously with order-flow and volatility-surface dynamics. 
These simplifications do not alter the core stability insight but highlight directions for structural refinement and empirical calibration. 

From a practical standpoint, the results provide quantitative guidance for risk management and hedging strategies. Traders and market participants can identify parameter regimes where feedback effects are likely to dominate, allowing for informed adjustments to position sizes, gamma exposure, or liquidity provisioning. For academic purposes, the model offers a tractable framework for exploring nonlinear feedback in derivative markets, with clear extensions possible for stochastic volatility, time-dependent hedging, or multi-asset interactions. 
Future work could be further extended by empirically fitting $I()$ from real market data and adjust parameters to provide guidelines for risk managements.

\section{Conclusion}
\label{sec:conclusion}
In this study, we developed a quantitative framework to analyze market-maker feedback dynamics and amplification effects in the presence of gamma exposure, hedging sensitivity, and exogenous price shocks. A formal gamma-squeeze condition that combines classical delta-hedge mechanics with beta sensitivity has been derived. This quantitative framework, 
through a combination of analytical derivations and numerical simulations, leads to a finding that low-beta, high gamma-exposure stocks are shown to be most vulnerable to endogenous volatility amplification.
These findings provide both theoretical insights and practical guidance for risk management, offering a structured approach to anticipate and mitigate destabilizing feedback effects in markets with significant derivative activity. The framework is suitable for simulation studies and empirical validation, offering both rigorous derivation and intuitive understanding. 

\section*{Appendix A. Summary of Symbols and Definitions}
\begin{table}[H]
\centering
\renewcommand{\arraystretch}{1.2}
\begin{tabular}{l p{0.72\textwidth}}
\hline
\textbf{Symbol} & \textbf{Definition / Economic Interpretation} \\
\hline
$S_t$ & Stock price at discrete time $t$. \\
$\Delta S_t$ & Incremental price change, $\Delta S_t = S_t - S_{t-1}$. \\
$\mu_t$ & Exogenous or semi-endogenous shock term; 
\\
$N_t$ & Active option position (option volume/exposure) decaying with cumulative movement. \\
$\bar{N}_t$ & Effective stochastic exposure: $\bar{N}_t = N_t + \nu_t$. \\
$\nu_t$ & Random incoming option exposure, 
\\
$\Gamma_t$ & Option gamma, measuring curvature of option delta with respect to price. \\
$\lambda_t$ & Hedging impact coefficient capturing price sensitivity to dealer flow. \\
$\beta$ & Stock beta, linking idiosyncratic shocks to market-level volatility. \\
$\sigma_m$ & Market-level volatility 
\\
$x_t$ & Beta-normalized relative surprise: $x_t = |\Delta S_t|/(S_t \beta\sigma_m) $
\\
$\phi(x_t)$ & Amplification function reflecting shock perception; typically $\phi = 1 + 2x_t$. \\
$I(y)$ & Saturation function limiting nonlinear amplification, $I(y)=\tanh(cy)$. \\
$\eta, \xi$ & Decay parameters controlling the rate and curvature of position contraction. \\
$\rho$ & Persistence of stochastic deviations in $\nu_t$. \\
$\sigma_N$ & Volatility scale of stochastic exposure noise. \\
$G$ & Total gamma exposure: $G = N_t \Gamma_t$. \\
$D_i$ & Stability denominator: $D_i = 1 - \lambda G \phi(x_i)$, threshold condition for squeeze onset. \\
$T$ & Simulation horizon (number of discrete time steps). \\
$S_0$ & Initial stock price. \\
\hline
\end{tabular}
\caption{Summary of principal symbols used in the gamma-feedback and stochastic recursive models. Only major variables necessary for model interpretation and simulation are listed.}
\end{table}

\section*{CRediT authorship contribution statement}
Haoying Dai: conceptualization, methodology, formal analysis, writing – original draft, visualization, supervision.

\section*{Declaration of competing interest}
The author declares that there are no known competing financial interests or personal relationships that could have influenced the work reported in this paper.

\section*{Acknowledgments}
The author acknowledges independent research support provided by Eight Sigma Research, a privately operated research initiative.

\section*{Data availability}
The simulation code and data that support the findings of this study are available from the corresponding author upon reasonable request.

\section*{Author biography}
\textbf{Haoying Dai} received the B.S. degree in Electronics from Southeast University, Nanjing, China, in 2015, and the Ph.D. degree in Electrical and Computer Engineering from the University of Maryland, College Park, MD, USA, in 2025, where he conducted his doctoral research at the Institute for Research in Electronics and Applied Physics (IREAP). His research interests include machine learning based on microwave photonic systems and chaos prediction using machine learning algorithms. He also studies the dynamics of complex systems, including financial systems.

\bibliographystyle{elsarticle-num-names}
\bibliography{gamma_beta_refs}

\end{document}